\documentstyle[twocolumn,epsf,prb,aps]{revtex}
\begin{document}
\preprint{IUCM95-039}
\draft
\twocolumn[\hsize\textwidth\columnwidth\hsize\csname
@twocolumnfalse\endcsname
\title{First Order Superconductor to Insulator Transition:
Evidence for a Supersolid Phase}
\author{Erik S.\ S\o rensen$^1$ and Eric Roddick$^2$}
\address{$^1$Department~of~Physics,
Indiana~University, Bloomington, IN~47405\\
$^2$Department~of~Physics,
The~Ohio~State~University, Columbus, OH~43210}
\date{\today}
\maketitle
\begin{abstract}
The superconductor to insulator transition in the
presence of long-range Coulomb interactions
is studied using Monte Carlo techniques.
At integer filling the transition is second
order. At half filling we find evidence of
a first order transition to a state with non-zero
$S_{\pi,\pi}$. No intervening supersolid
phase is observed. Off half-filling two separate transitions are
found. The transition from the superconducting to a supersolid phase
is second order followed by another second order transition between the
supersolid and an insulating phase.
\end{abstract}
\pacs{74.20.Mn, 05.30.Jp, 67.90.+z}
\vskip2pc]
If a phase possesses both diagonal long range order (DLRO), associated
with a structural ordering, as well as off diagonal long range order (ODLRO),
indicating phase coherence, it is referred to as a supersolid.
Initially Penrose and Onsager~\cite{penrose} showed that such a phase cannot
occur if each lattice site is always occupied. However, it was later realized
that if the occupancy is not an integer, and if the wave-function is properly
symmetrized, a supersolid phase can exist~\cite{leggett}. Analogous
phases in spin models were investigated and
shown to be possible~\cite{mefisher}.
Experimental
evidence for a supersolid phase in solid $^4$He seems at present
scant~\cite{meisel} although recent experiments have seen indications of
novel behavior~\cite{lengua} possibly related to a superhexatic
phase~\cite{mullen}. Exploiting a duality argument~\cite{fisherlee}
the analogue of a supersolid phase of vortex lines in type-II
superconductors has been
proposed~\cite{frey}. Recent numerical studies
have found evidence of supersolid phases in quantum lattice
models~\cite{eric,otterlo,scalettar}.
These studies indicate that interactions
longer range than on-site are necessary to stabilize a supersolid phase.
However, true long-range interactions were not considered, and the
order of the transition was not investigated numerically.

In this communication we study the phase diagram of the 2D
{\it charged} lattice
Bose gas at several different filling factors. This model should be
applicable to short coherence length superconductors where the individual
Cooper pairs have formed prior to the
superconducting transition~\cite{nextpaper}.
At integer fillings we find a second order transition with 3D $XY$-like
exponents in agreement with predictions by Ma~\cite{ma}.
At half-filling we find a single {\it first order} transition from the
superconducting phase into an insulating phase with checkerboard charge
ordering. No supersolid phase is found. Slightly below half-filling
we find a second order transition from the superconducting
phase into a supersolid phase with ODLRO and charge ordering
followed by another second order transition into an insulating phase.
A first order supersolid transition has recently been observed in a
soft sphere model~\cite{pomeau}.
A detailed account of our results will
be published elsewhere~\cite{us}.

Our starting point is the boson Hubbard model~\cite{fisher89c}
\begin{eqnarray}
H&=&\frac{U}{2}\sum_{\bf r}\hat n_{\bf r}^2-
\sum_{\bf r}(\mu-zt)\hat n_{\bf r}
-t\sum_{\langle {\bf r},{\bf r'}\rangle }
(\hat \Phi^\dagger_{\bf r}\hat \Phi_{\bf r'}+
\hat \Phi_{\bf r}\hat \Phi^\dagger_{\bf r'})
\nonumber\\
&+&\frac{{e^{\ast}}^2}{2}
\sum_{\bf r, r'}\hat n_{\bf r} G(r - r')\hat n_{\bf r'}\ .
\label{eq:bosonhubbard}
\end{eqnarray}
Here $U$ is the on-site repulsion,
$\mu$ is the chemical potential, z the number of nearest neighbors,
and $\hat n_{\bf r}=\hat \Phi^\dagger_{\bf r}\hat \Phi_{\bf r}$
is the number operator on site ${\bf r}$, with $\Phi^\dagger_{\bf r},
\Phi_{\bf r}$ boson creation and annihilation operators.
The hopping strength is given by $t$
and $G$ is the $1/r$ Coulomb potential modified for the periodic
lattice. We shall use either an Ewald or lattice Greens function for
$G$ with ${e^{\ast}}^2=1/2$.
This model can be mapped onto a link-current
model~\cite{fisherlee,nextpaper,cha,JKKN}
defined in terms of integer currents ${\bf J} = (J^x,J^y,J^{\tau})$
on the links of the (2+1)D lattice
where ${\bf r} = (x,y)$ and $\tau$ is the imaginary time which has been
discretized. Note that
$J^{\tau}$ is just the particle density.
The integer-link variables
$(J^x,J^y,J^{\tau})$ take integer values from
$-\infty$ to $\infty$, subject to the divergence-free constraint
given by the continuity equation, ${\bf \nabla} \cdot{\bf J} = 0$.
If we include long range Coulomb interactions
we can write down a purely {\it real\/}
action in this representation~\cite{nextpaper}
\begin{eqnarray}
S&=&\frac{1}{K}
\sum_{{\bf r},\tau}
\Bigl\{ \frac{1}{2}{\bf J}^2 ({\bf r}, \tau)
- \mu J^{\tau}({\bf r},\tau) \Bigr\}\nonumber\\
&+& \frac{{e^{\ast}}^2}{2K}
\sum_{{\bf r},{\bf r}^{\prime},\tau}
[J^{\tau}({\bf r},\tau)-n_0] G({\bf r}-{\bf r}^{\prime})
[J^{\tau}({\bf r}^{\prime},\tau)-n_0] .
\label{eq:ourH}
\end{eqnarray}
The calculation is performed in the canonical ensemble.
Hence, the term involving $\mu$ is
a constant and is only included for computational convenience.
At integer filling in the absence of long-range interactions this
model is in the same universality class as the 3D $XY$ model and
has an {\it inverted} $XY$ transition~\cite{JKKN,oldsc}. Including
long-range interactions previous studies have found that the transition
can be either first or second order~\cite{old2} at integer filling.

Having arrived at an equivalent classical model we can employ standard
Monte Carlo methods to study the insulating transition by letting $K$
take the place of a ``temperature". See Ref.~\onlinecite{nextpaper}
for a detailed description. We first locate the approximate
position of the critical point, $K_c$, with low statistics calculations.
Then the histogram Monte Carlo method~\cite{ferrenberg} is used to obtain
high precision results for a range of $K$ close to $K_c$. For each system
size $3-10\times10^4$ Monte
Carlo sweeps (MCS) of the entire lattice are discarded for
equilibration followed by
$3-10\times10^4$ MCS for calculations with a measurement
every 10 MCS.
In order to calculate the error-bars we repeat the entire calculation
50 times for each lattice size.
Two replicas with different random number sequences
are run in parallel allowing overlaps between them to be calculated.
Techniques
used in the study of spin glasses then allow us to check that we
are calculating equilibrium quantities~\cite{bhatt}. Thus we use
in total $6-20\times 10^6$ MCS to build the final probability distribution
for use in the histograms.

In order to study the structural ordering we define the structure
factor, $S_{\bf k}$, as follows:
\begin{equation}
S_{{\bf k}}= \frac{1}{L^4 L_{\tau}}
\left\langle
\sum_\tau
\left|\sum_{\bf r}
e^{i{\bf k}\cdot {\bf r}}
J^\tau_{({\bf r},\tau)}\right|^2
\right\rangle.
\end{equation}
Note that $S_{\bf k}$ is positive definite. At half filling we
expect the dominant ordering to be associated with the
wave-vector $(\pi,\pi)$ corresponding to a checkerboard ordering with basis
vectors $(a,a)$ and $(a,-a)$.
The onset of the superconducting phase will be associated
with a non-zero stiffness~\cite{nextpaper} defined as
\begin{equation}
\rho = \frac{1}{L^d L_\tau}
\left\langle\left(\sum_{({\bf r},\tau)}
J^x_{({\bf r},\tau)}\right)^2\right\rangle  \ .
\end{equation}
Note that this stiffness corresponds to a non-zero Drude weight~\cite{zhang}.
Invoking finite-size scaling we arrive at the following finite-size
scaling form for the stiffness
\begin{equation}
\rho = {1 \over L^{d+z - 2}} \bar\rho \left(
L^{1/\nu} \delta,\ {L_\tau \over L^z} \right).
\label{eq:rfinscale}
\end{equation}
Here $z$ is the dynamical critical exponent defined through
the relation $\xi_\tau\sim\xi^z$, and $\delta=K-K_c$ is the distance
to the critical point, $K_c$.
Assuming that the transition is second order,
we see From Eq.~(\ref{eq:rfinscale})
that if the aspect ratio, $c={L_\tau /L^z}$, is kept
constant $L^{d+z - 2}\rho$ should be independent of $L$ at the critical
point. In the presence of long-range Coulomb interactions
{}Fisher~et~al~\cite{FGG} has argued that $z=1$ and we shall in
the following take that as our starting assumption.

{\it Integer filling.}
{}For a filled lattice we do not expect any non-trivial charge ordering
and there should only be a single transition between the superconductor
and a Mott insulating phase, governed by a dynamical critical exponent
$z=1$~\cite{FGG}.
We therefore locate the critical point by calculating
$L\rho$ using cubic samples. A well defined crossing is found at
$K_c=0.289(2)$~\cite{us}. Using the
finite size scaling form $L\rho=\tilde\rho(\delta L^{1/\nu})$ a scaling
plot of $L\rho$ can be used to determine $\nu$.
We find $\nu\simeq 0.67$~\cite{us}.
We also calculate the correlation functions both
in the time and space direction.
This yields an independent calculation of the dynamical critical exponent, $z$,
and the anomalous dimension $\eta$~\cite{nextpaper}.
By fitting the correlation functions at
the critical point to power law forms we find $z=1.0\pm 0.1,\ \eta = -0.15\pm
0.2$, in reasonable agreement with the prediction by Ma~\cite{ma} that the
exponents should remain 3D-$XY$ like due to the irrelevance of Coulomb
interactions for this filling factor.
However, critical amplitudes could possibly be affected by the
Coulomb interactions and a priori we cannot expect to find the same value
of the universal resistivity~\cite{FGG} at the critical point
as was found for
the 3D-$XY$ model~\cite{cha}.
If we calculate the frequency dependent stiffness:
\begin{equation}
\rho(i\omega_n) =
\frac{1}{L^2L_{\tau}} \left\langle \left| \sum_{({\bf r},\tau)}
e^{i\omega_n\tau}
J^x_{({\bf r},\tau)} \right|^2 \right\rangle,
\label{eq:rhon}
\end{equation}
the conductivity can be calculated from the relation $\sigma(i\omega_n)=2\pi
\sigma_Q \rho(i\omega_n)/\omega_n$. Here the quantum of conductance is
$\sigma_Q\equiv(2e)^2/h\equiv R_Q^{-1}$
and $\omega_n=2\pi n/L_\tau$ is the $n$th Matsubara
frequency. With a $1/L$ finite size correction it is possible to collapse
all the data for different lattice sizes onto a single curve at $K_c$
and we find $R_\Box^\ast/R_Q=1.5\pm0.4$.

\begin{figure}
\centering
\epsfysize=7.5 cm
\leavevmode
\epsffile{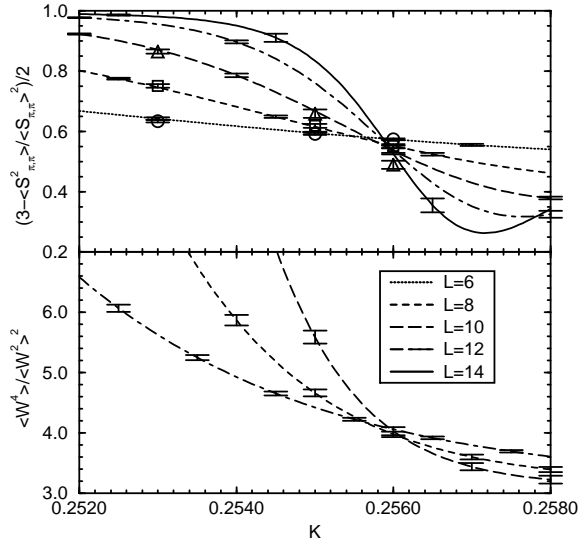}
\caption{The cumulant ratio of the structure factor, $S_{\pi,\pi}$, and
winding number, $W$, as a function
of $K$ at half-filling.
The lines indicate results obtained using
the histogram Monte Carlo method for different lattice sizes.
The circles, squares and triangles denote
seperate calculations with lesser precision for lattices of size $L=6,8,10$,
respectively.
}
\label{fig:bball}
\end{figure}

\begin{figure}
\centering
\epsfysize=7.5 cm
\leavevmode
\epsffile{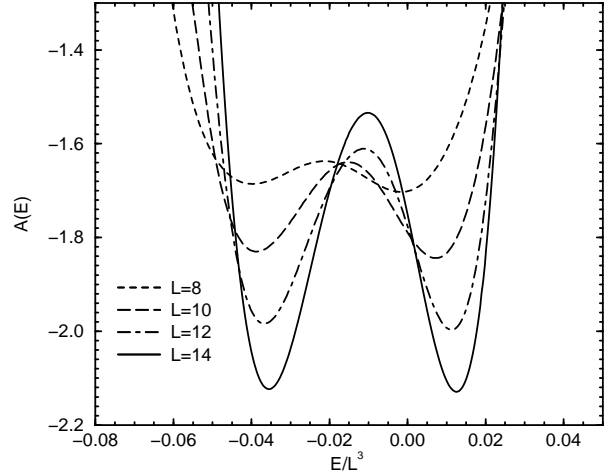}
\caption{$A(E)=-\ln P(E)$ as a function of
$E/L^3$ for the lattice sizes $L=8,10,12,14$
at half-filling.
$P(E)$ was calculated at $K_c(L)=0.25345,
0.25465, 0.25504, .25548$, respectively.
The relative height of the central peak increases with $L$.
}
\label{fig:lee}
\end{figure}

{\it Half-integer filling.}
We now turn to a discussion of the half filled case.
In this case we presumably also have $z=1$~\cite{FGG,scalettar},
however, none of the other exponents are known.
In order to locate the structural transition associated with the
assumed $(\pi,\pi)$ ordering we calculate a cumulant
ratio of the structure factor $S_{\pi,\pi}$ defined as:
$g(K)=(3-<S^2_{\pi,\pi}>/<S_{\pi,\pi}>^2)/2$.
With this definition $g(K)$ should approach 1 in the ordered phase and
0 in the disordered phase. At the critical point $g(K)$ should
be independent of $L$ and curves representing $g(K)$ for different
lattice sizes should therefore intersect close to $K_c$ {\it independent} of
the order of the transition~\cite{hermann}. Our results are shown
in Fig.~\ref{fig:bball} for the lattice sizes $L=6,8,10,12,14$.
The curves for different $L$ intersect at
$K_c=0.2559(4)$. However, above $K_c$ the curves for the
larger $L$ develops a minimum and starts to increase.
This behavior is usually associated with a first order phase
transition~\cite{hermann}. A plot of the probability distribution,
$P(E)$ for the larger lattices shows a double peak structure
close to $K_c=0.2559(4)$.
{}Following Lee and Kosterlitz~\cite{lee}
we consider $A(E,L)\equiv -\ln P(E,L)$. Here
$P(E,L)$ is calculated at $K_c(L)$, the coupling at which
the probability distribution has two peaks of equal heights
at $E=E_1,E_2$ surrounding a minimum at $E_m$. One can
then define a bulk free-energy barrier between the ordered
and disordered states as $\Delta F(L)=A(E_m,L)-A(E_i,L)$.
In Fig.~\ref{fig:lee} we show results for $A(E)$ for
$L=8,10,12,14$. As a function of $L$, $\Delta F(L)$
increases monotonically. However, since $\Delta F(L)$
has not approached the asymptotic form
$\Delta F(L)\sim L^{d+z-1}$~\cite{lee}
we expect that we in all cases have $L<\xi$.
$S_{\pi,\pi}$ is significantly different in the two minima.

In Fig.~\ref{fig:bball} we show results for the cumulant ratio associated
with the winding number and hence the onset of superfluidity.
This quantity as well as $L\rho$ shows that the superfluid
order parameter disappears at the same $K_c$ where the first
order structural transition takes place. We therefore conclude that the
action Eq.~(\ref{eq:ourH}) has a {\it single} first order phase transition
at half-filling. We believe that this implies that the underlying
quantum Hamiltonian therefore also has a first-order transition,
although it is possible to argue that neglected irrelevant terms
in Eq.~(\ref{eq:ourH}) could cause the transition to be second order
for the quantum problem.
However, we have explicitly checked that
the transition remains first order
for both the Ewald and lattice Greens function form
of the potential.
The above scenaro is in
good agreement with recent work by Batrouni {\it et al.},
and Scalettar {\it et al.}~\onlinecite{scalettar}.

\begin{figure}
\centering
\epsfysize=7.5 cm
\leavevmode
\epsffile{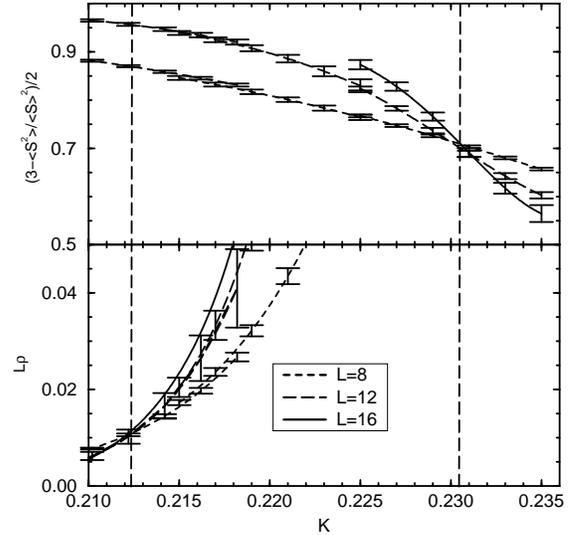}
\caption{Results for a filling factor of 7/16. Shown are the
the cumulant ratio for the structure factor, and $L\rho$
as a function of $K$.
}
\label{fig:offhalf}
\end{figure}

{\it Off half-filling.}
As the range of the interaction is increased a particulary rich phase
diagram emerges for the boson Hubbard model~\cite{bruder}.
At half-filling the insulating phase at small $K$ corresponds
to a checkerboard charge ordering with unit cell twice as large as the
underlying lattice.
Due to the limited lattice sizes that it is possible to study
numerically, the most convenient filling factor away from half-filling is
7/16. At this filling factor we find, in the limit $K\to 0$,
an insulating phase
with charge ordering corresponding to vacancies in a $(0,4a)$, $(4a,2a)$,
$(4a,-2a)$ triangular lattice superimposed on the half-filled
$(\pi,\pi)$ ordering~\cite{us}.
Analogous orderings can be found at other
filling factors.
This charge ordering is commensurate with the lattice sizes
$L=8,16$.
Thus, at this filling factor we can use the $S_{\pi,\pi}$ structure factor
to monitor any charge ordering even though the actual ground-state
corresponds to ordering at more than one wave-vector.
Scalettar et al~\cite{scalettar} have shown that $z=1$ at the
superfluid-supersolid transition. Due to the long range interactions we
expect $z=1$ also at the
supersolid-solid transition as opposed to the short-range
case~\cite{scalettar}. In our simulations we therefore use $z=1$ throughout.
Our results are shown in Fig.~\ref{fig:offhalf}.
Two separate transitions are visible at $K_{c1}=0.2125(15)$ and
$K_{c2}=0.2305(15)$ surrounding a supersolid phase.
At the transition between the superconducting and supersolid phase
we find
$1/\nu\sim 1.8\pm.3$.
The transition between the supersolid and insulating phase
is tentatively characterized by the exponent
$1/\nu\sim 1.1\pm0.3$, but a detailed analysis is hindered by the fact
that the $L=12$ system is not commensurate with the assumed
ground-state.
We have also calculated the structure factors $S_{\pi,0}$ and
$S_{0,\pi}$ corresponding to collinear ordering. In none of the above
cases have we observed charge ordering at these wave vectors.

In conclusion, we have found a {\em first-order} superconductor to
insulator transition at half-filling.
Off half filling we find evidence for a supersolid phase surrounded
by two second order phase transitions.

We gratefully acknowledge many helpful discussions with
S.~M.~Girvin, A.~H.~MacDonald,
D.~Stroud,  R.~\v{S}\'a\v{s}ik, and A.~P.~Young,
and support from DOE grant DE-FG02-90ER45427, through the Midwest
Superconductivity Consortium at Purdue University.
ESS is also supported in part by NSF grant number NSF DMR-9416906,
and ER by NSF DMR94-02131 and NSF DMR90-20994.

\end{document}